# Thermal Transport for Probing Quantum Materials

Mingda Li[1] and Gang Chen[2]
[1]Department of Nuclear Science and Engineering, MIT, Cambridge, MA 02139
[2]Department of Mechanical Engineering, MIT, Cambridge, MA 02139

**Abstract**

Thermal transport is less appreciated in probing quantum materials in comparison to electrical transport. This article aims to show the pivotal role that thermal transport may play in understanding quantum materials: the longitudinal thermal transport reflects the itinerant quasiparticles even in an electrical insulating phase, while the transverse thermal transport such as thermal Hall and Nernst effect are tightly linked to nontrivial topology. We discuss three types of examples: quantum spin liquids where thermal transport identifies its existence, superconductors where thermal transport reveals the superconducting gap structure, and topological Weyl semimetals where anomalous Nernst effect is a consequence of nontrivial Berry curvature. We conclude with an outlook of the unique insights thermal transport may offer to probe a much broader category of quantum phenomena.

**Introduction**

Quantum materials encompass a broad category of condensed matter phases that emergent quantum phenomena play a decisive role in materials properties of interest (*1*), such as conventional and unconventional superconductors (*2-6*), band topology such as quantum Hall families (*7-15*), topological insulators (*16-18*) and topological semimetals (*19-22*), particularly Dirac and Weyl semimetals (*23-26*), frustrated magnetism and multiferroics (*27-33*), graphene and low-dimensional van der Waals heterostructures (*34-40*), among others. The manifestation of quantum effects usually needs low temperatures with suppressed thermal fluctuation. While electrical transport is often used to probe quantum phenomena, thermal transport seems to be less appreciated as it will unavoidably introduce some finite temperature effects. Then, a natural question to ask: why thermal transport in quantum materials?

To address this question, it might be worthwhile to point out that although we can always perform thermal transport measurement in a quantum material, it may not necessarily tell more information beyond electrical transport measurement. In this perspective, we limit the scope of this paper to where thermal transport can provide unique insights, with an emphasis on the emergent phenomena originated from correlation effect and topology. The total thermal conductivity $k_{tot}$ of a material can be written as the sum of contributions from electrons ($k_e$), phonons ($k_{ph}$) and other carriers ($k_{other}$)

$$k_{tot} = k_e + k_{ph} + k_{other} \tag{1}$$

For a large category of materials, the electronic contribution $k_e$ is related to electrical conductivity $\sigma$ via the Wiedemann-Franz law (*41*)

$$\frac{k_e}{\sigma T} = L \tag{2}$$

where the Lorenz number $L \equiv \frac{\pi^2}{3}\left(\frac{k_B}{e}\right)^2 = 2.44 \times 10^{-8}\,\text{W}\Omega/\text{K}^2$ is a universal constant in metals and depends on carrier density in semiconductors. A large deviation of the Lorenz number from the universal value provides a clue to quantum transport, as is observed in the metal-insulator transition (*42*). At low-temperature, phonon thermal conductivity $k_{ph}$ can generally be well described by the Debye's model, i.e., $k_{ph} \propto T^3$ for isotropic material. For an electrically insulating material with $k_e = 0$, the measurement of total thermal conductivity $k_{tot}$ becomes suitable to probe other itinerant quasiparticles $k_{other}$ beyond phonon contribution. As we will see shortly, this is the case why thermal transport plays a crucial role in quantum spin liquids (QSLs).

The above argument largely applies to longitudinal transport. Transverse transport as represented by the Hall effect has been extensively used to understand charge transport (**Figure 1b**). A longitudinal temperature gradient $\nabla T$ can also generate transverse effects. Examples are the Nernst effect which is the transverse voltage $\nabla V_H$ and the Regi-Leduc effect which is the transverse temperature gradient $\nabla T_H$, both generated when there is a longitudinal temperature gradient $\nabla T$, usually under an external magnetic field $H$ (*43*). While the Regi-Leduc effect typically is constrained to electrically conducting materials, transverse temperature gradients can also develop in as long as the quasiparticles are responsive to external magnetic field (**Figure 1b**), and is generally called thermal Hall effect. In a more intriguing case, just like the integer quantum Hall effect with topological protected chiral edge electrical current (*14*), or quantum spin Hall effect with topologically protected edge spin current (*11*), the thermal Hall measurements can also feature the existence of nontrivial topology (**Figure 1c**). Indeed, just like longitudinal thermal transport, which is sensitive to charge-neutral itinerant quasiparticles, here the thermal Hall effects can also be used to detect edge states carrying neither electrical charge nor spin, such as charge-neutral Majorana fermions in QSLs (*44, 45*) and topological superconductors (*46*), as well as fractionally charged and charge-neutral edge quasiparticles in fractional quantum Hall systems (*47, 48*). The Nernst effect has been employed to reveal Cooper pair fluctuations and magnetic vortices in superconductors (*49*), and more recently becomes a hallmark of the nontrivial Berry curvature in topological Weyl semimetals (*50-52*).

**Nature of quantum spin liquids**

QSLs are a novel class of condensed matter phase where a strong quantum fluctuation of spin prevents the formation of any long-range magnetic ordering even at $T$=0 K (*30*). In the past decade, QSLs have attracted significant research interest, due to the a bsence of long-range spin correlation but simultaneous existence of long-range entanglement (which is in sharp contrast to a classical thermally driven disordered spin states), the exotic fractionalized, fermionic elementary quasiparticles, aka the spinons (which is in sharp contrast to Bosonic magnons in ordered magnet), the manifestation of topological ordering of the ground state degeneracy with potential quantum information processing applications, and the link to demystify unconventional superconductivity (*53, 54*). A small local spin $S=1/2$ is usually favored to allow for large quantum effect, and a key ingredient to form QSL is a means that can generate large spin frustration.

Until today, there are at least two categories of means that can realize large spin frustration: geometrical frustration and exchange frustration (**Figure 2a**). Geometrical frustration originates from a crystal lattice structure that un-favors spin alignment and inclines to have large ground-state degeneracy, such as triangular and kagome lattices in 2D, and hyper-kagome and pyrochlore lattices in 3D. In a 2D triangular lattice, the prototypical QSL candidate materials include the layered organic compound *κ*-(BEDT-TTF)$_2$Cu$_2$(CN)$_3$ (*55*) and EtMe$_3$Sb[Pd(dmit)$_2$]$_2$ (*56*), and the inorganic compounds quasi-2D YbMgGaO$_4$ (*57, 58*) and YbZnGaO$_4$. As to the 2D kagome lattice, the ZnCu$_3$(OH)$_6$Cl$_2$, belonging to the Herbertsmithite structure consisting of 2D kagome lattice of Cu$^{2+}$ ions with spin 1/2, has attracted considerable interest (*59, 60*). For the 3D lattices, geometrical frustration can be established by forming networks of 2D frustration units, such as triangular networks in pyrochlore and hyper-kagome lattices; the pyrochlore lattice hosts the spin-ice states (*32*), which makes the search for QSL state from spin ice viable (*30, 61*), while the hyperkagome lattice compounds such as Na$_4$Ir$_3$O$_8$ can also host potential QSL state (*62*).

With extensive research in geometrically frustrated QSL candidate materials, there is still a substantial difficulty in that the theoretical deduction of a QSL ground state from a given frustrated lattice structure has to go through a series of approximations and intuitions. A completely different avenue to achieve QSL is proposed by Kitaev (*63*), which utilizes the non-commutativity of spin operators along different directions to generate frustration, moreover is exactly solvable with a QSL ground state. The candidate materials include the Na$_2$IrO$_3$ (*64*) and *α*-RuCl$_3$ (*44, 65-67*) in a 2D honeycomb lattice, and the Li$_2$IrO$_3$ in 3D variants of honeycomb lattices (*68*), etc.

Given the many excellent reviews available (*30, 53, 54, 69-71*) and the rapidly growing literature even restricted to thermal properties (*44, 65, 72-84*), it is hard to summarize even a small portion of the active field of QSL; however, the critical importance of thermal transport is not hard to appreciate: the hallmark signature of a QSL state is a fractionalized gapless elementary excitation, called the spinon, which can be directly observed from thermal transport (**Figure 2b**). In the current context, "gap" means the energy gap between the ground state and the 1$^{st}$ excited state in the many-body materials system. In conventional ordered magnet with elementary excitation of magnons (**Figure 2c**), the magnon contributed thermal conductivity $k_{mgn}$ can be written in terms of an Arrhenius form

$$k_{mgn} \propto T \exp(-\Delta_M / k_B T) \tag{3}$$

where $\Delta_M$ is the magnetic energy gap magnitude. For Zeeman-split gap, we have $\Delta_M = g\mu_B B$, in which $g$ is the $g$-factor and $B$ is the magnetic field, $\mu_B$ is the Bohr magneton. In the gapless spinon excitation case, the corresponding thermal conductivity $k_{spn}$ has a simple linear relation with $T$

$$k_{spn} \propto T \tag{4}$$

Since phonon-contributed thermal conductivity $k_{ph}$ largely obeys the Debye's model $k_{ph} \propto T^3$, in order to identify that whether a candidate insulating material ($k_e = 0$) may contain spinon contribution to thermal transport, we can plot $k_{tot}/T$ as a function of *T*. In a candidate material with conventional magnetic ordering, there is no residual thermal conductivity in the $T \rightarrow 0$ limit, i.e., $k_{tot}/T$ will pass through the origin in the $k_{tot}/T$ vs *T* plot (**Figure 2d**); on the contrary, in a QSL candidate, we have

$$\frac{k_{tot}}{T} = \underset{\text{spinon}}{a} + \underset{\text{phonon}}{bT^2} \tag{5}$$

where $k_{tot}/T$ will intercept at a finite value at *T*=0 K (**Figure 2b**). The non-vanishing linear residual term ($a \neq 0$) in the $T \rightarrow 0$ limit can thus serve as a strong indication for the existence of spinon excitations from QSL state.

Despite the deceptively simple discriminator of QSL state using Eq. (5), the reality is much more complicated with constant controversy exists. That is why the term "QSL candidate" is preferred in many contexts. In fact, during the development of QSLs, there are many QSL candidates later found not to be true QSL state, such as kagome $Cu_3V_2O_7(OH)_2 \cdot 2H_2O$, triangular $Cs_2CuCl_4$, YbZnGaO$_4$(*85*), and most recently triangular $EtMe_3Sb[Pd(dmit)_2]_2$, which is shocking to some extent (**Figure 2 d-g**). On the one hand, $EtMe_3Sb[Pd(dmit)_2]_2$ has a large exchange coupling constant *J*~220-250K (as a comparison, YbMgGaO$_4$ only has *J*~1.5K), indicating that even at the same absolute temperature value, the condition *T*<<*J*, which is the suitable range to identify QSL, is much easier to meet. On the other hand, thermal transport has clearly demonstrated a large linear residual term $a \neq 0$ in $EtMe_3Sb[Pd(dmit)_2]_2$ a decade ago (**Figure 2d**) (*56*), and had been studied intensely since then. However, two very recent thermal transport experiments on the material reported a measured $a = 0$ (**Figure 2e, f**) (*86, 87*), excluding the possibility of it being a QSL material. The crucial role of thermal transport to QSL is clear from these studies. It is also worthwhile mentioning that heat capacity *C* measurement can also show a QSL-like behavior with apparent $a \neq 0$, (**Figure 2g**). However, since *C* measures both itinerant quasiparticles and other contributions such as Schottky anomaly from local ions (*88*) while *k* only measures itinerant quasiparticles, in the context of QSL where itinerant spinons is of interest, *k* is generally considered more reliable over *C*.

In parallel to the longitudinal thermal transport discussed above, the QSL state can also be linked to the existence of thermal Hall conductivity. In particular, when nontrivial Berry curvature exists, such as in a kagome lattice magnet (*89*), or by considering a Fermi liquid of spinons (*90*), a thermal Hall effect can be developed. If the longitudinal spinon contributed thermal conductivity is written as $k_{spn}^{xx} = aT$, then in a generic "Fermi liquid" of spinons in QSL, the thermal Hall conductivity $k_{spn}^{xy} \approx k_{spn}^{xx} \times (\omega_c \tau)$, where $\omega_c$ is the effective spinon cyclotron frequency, $\tau$ is the spinon lifetime (*90*). Furthermore, in a 2D Kitaev QSL, where the 1D edge current composed of chiral Majorana fermions exists (**Figure 3a**), the thermal Hall conductivity approaches to a quantized universal value (*44*)

$$\frac{k^{xy}}{T} = q \times \frac{\pi}{6} \frac{k_B^2}{\hbar} \qquad (6)$$

where the fractionalized $q = 1/2$ is the signature of itinerant chiral Majorana fermions moving along the sample edge. Recent measurements on α-RuCl₃ are consistent with this picture, suggesting the observation of chiral Majorana fermions (**Figure 3b**).

**Superconducting energy gap**

In a superconductor, given the vanishing electrical resistivity $\rho_e = 0$ below the superconducting critical temperature $T_c$, the electron transport measurements below $T_c$ does not contain much useful information. Furthermore, the Cooper pairs in the superconducting state do not carry heat. However, even the Cooper pairs do not carry entropy and do not contribute to thermal transport, thermal transport is still valuable by probing delocalized low-energy quasiparticles (*91*). Hence, thermal transport at $T < T_c$ has been extensively studied, including a wide variety of unconventional superconductors, such as experimental (*92-98*) and theoretical (*99-105*) studies in high-$T_c$ cuprates, multi-band superconductor MgB₂ (*106-108*), heavy fermion superconductors (*109-111*), and more recently iron-based superconductors (*112-116*). At the superconducting phase transition, specific heat $C_p$ develops a certain level of upturn. Since thermal conductivity $k$ from kinetic theory without distinguishing itinerant from localized excitations gives $k = \frac{1}{3} C_p v l$, where $v$ is carrier velocity and $l$ is the mean free path (*117*), it is expected that the thermal conductivity may show observable signatures below $T_c$.

The most useful aspect of thermal transport to superconductivity probably lies in the determination of the symmetry of the superconducting order parameter (*91, 118, 119*), which is proportional to the superconducting gap function $\Delta(\mathbf{k})$. Here the gap means the energy gap of the superconducting quasiparticles. An isotropic *s*-wave superconductor, representing the prototype of many conventional superconductors, can have a simple, constant gap value $\Delta_0$ across the entire Brillouin zone:

$$\Delta_s(\mathbf{k}) = \Delta_0 \tag{7}$$

indicating a fully gapped spectra for superconducting quasiparticles. On the other hand, in a *d*-wave superconductor, such as a $d_{x^2-y^2}$ superconductor, which is the prototype for high $T_c$ cuprates, the gap can be written as (*2*)

$$\Delta_{d_{x^2-y^2}}(\mathbf{k}) = \Delta_0(\cos k_x - \cos k_y) \tag{8}$$

where the gapless $\Delta_{d_{x^2-y^2}}(\mathbf{k}) = 0$ state can be reached at $k_x = \pm k_y$, called nodal superconductors with the presence of gapless nodes (**Figure 4a**). Here, the gapless **k**-points $k_x = \pm k_y$ form a line-shape in the Brillouin zone, called nodal lines. Similar to the case of QSL where gapless state has a finite linear residual term $a \neq 0$ while gapped state has $a = 0$, if we plot $k/T$ vs $T$ at $T \to 0$ limit, it can tell that whether the system is a gapped *s*-wave superconductor ($a = 0$) or a *d*-wave superconductor containing gapless nodes ($a \neq 0$ in Eq. 5), as shown in **Figure 4b**. The similarity of the linear-*T*-dependent term in both QSLs and superconductors can be understood from the Wiedemann-Franz law, but applied to spinon Fermi surface and nodes in superconductors, respectively (*90, 101, 120*), where electrical conductivities become temperature independent constant. Based on this spirit, it has been shown that superconducting $RbFe_2As_2$ and $Ba(Fe_{1-x}Co_x)_2As_2$ contain gapless nodes (*113, 115*). It is also particularly interesting to mention that the intercept of $k/T$ at $T = 0\,\mathrm{K}$ for a nodal superconductor obeys a universal relation, that (*101*)

$$\frac{k}{T} = \frac{\pi^2}{3} N(E_F) v_F^2 \frac{a\hbar}{2\mu\Delta_0} \tag{9}$$

in which $N(E_F)$ is the density of states at Fermi level $E_F$, $v_F$ is the Fermi velocity at the gapless node, $\mu$ is a constant (for *d*-wave superconductor $\mu = 2$) and *a* is a constant order of unity. As a result, the universal thermal conductivity in nodal superconductors offers a valuable microscopic lens on the electronic properties.

**Nernst hallmark of Berry curvature**

Nernst effect has been studied in metals and superconductors (*49*) (**Figure 5a**), and can develop a large signal across a Lifshitz transition where Fermi surface experiences a change of topology (*121, 122*). Similar to the anomalous Hall effect, even without external magnetic field, the Nernst effect can still emerge, called anomalous Nernst effect (ANE). ANE has been extensively studied in ferromagnetic metals, however, given the direct link between ANE and the Berry curvature, it has gained significant recent theoretical (*123-131*) and experimental (*50-52, 132-138*) attention in the field of topological Weyl semimetals (**Figure 5b**). A Weyl semimetal carries linear-dispersive low-energy quasiparticles of Weyl fermions with definitive chirality, which is directly linked to the Berry curvature singularity. The Berry curvature $\Omega(\mathbf{k})$ is the magnetic field in **k**-space. A simple way of seeing it is the symmetry between the simplified semi-classical electron equation of motion of coordinate **r** and the momentum **p**, that (*124*)

$$\hbar \dot{\mathbf{r}} = \nabla_{\mathbf{k}} \varepsilon(\mathbf{k}) + \dot{\mathbf{p}} \times \Omega(\mathbf{k})$$
$$\frac{1}{e}\dot{\mathbf{p}} = -\nabla_{\mathbf{r}} V(\mathbf{r}) + \dot{\mathbf{r}} \times \mathbf{B}(\mathbf{r}) \tag{10}$$

from which we can see immediately that even $\Omega(\mathbf{k})$ is defined in **k**-space, it has an observable influence on particle motion in real space. As a result, a finite Berry curvature along *z*-direction can lead to transverse anomalous Hall conductivity $\sigma_{xy}^{A}$ and ANE related thermoelectric tensor $\alpha_{xy}^{A}$, that (*51*)

$$\sigma_{xy}^{A} = \frac{e^2}{\hbar} \sum_{n} \int \frac{d^3\mathbf{k}}{(2\pi)^3} \Omega_{n,z}(\mathbf{k}) f_{n,\mathbf{k}}$$
$$\alpha_{xy}^{A} = \frac{k_B e}{\hbar} \sum_{n} \int \frac{d^3\mathbf{k}}{(2\pi)^3} \Omega_{n,z}(\mathbf{k}) s_{n,\mathbf{k}} \tag{11}$$

where *n* is the band index, $f_{n,\mathbf{k}}$ is the Fermi-Dirac distribution function, and $s_{n,\mathbf{k}} = -f_{n,\mathbf{k}} \ln f_{n,\mathbf{k}} - (1-f_{n,\mathbf{k}}) \ln(1-f_{n,\mathbf{k}})$ is the occupational entropy. Since fully occupied band ($f_{n,\mathbf{k}} = 1$) and empty band ($f_{n,\mathbf{k}} = 0$) do not carry entropy ($s_{n,\mathbf{k}} = 0$), we can see that finite $\alpha_{xy}^A$ can be acquired near chemical potential with partially filled band ($0 < f_{n,\mathbf{k}} < 1, s_{n,\mathbf{k}} \neq 0$), and $\alpha_{xy}^A$ will be increased with large Berry curvature projection $\Omega_{n,z}(\mathbf{k})$.

In topological Weyl semimetals, the Weyl nodes are monopole-like singularities of Berry curvature, and the paired left-handed and right-handed Weyl nodes emerge as source and sink of Berry curvature. In this light, large ANE, along with large transverse thermopower $S_{xy} = \dfrac{-\nabla_y V}{-\nabla_x T} = \dfrac{\alpha_{xy}\sigma_{xx} - \alpha_{xx}\sigma_{xy}}{\sigma_{xx}^2 + \sigma_{xy}^2}$ can emerge when Berry curvature is large (**Figure 6a**, the red blurred region near Weyl nodes in cases (i)-(iii)). In particular, when the chemical potential is located right at the Weyl node where Berry curvature diverges, divergent $S_{xy}$ can be obtained (**Figure 6a,** case (iii)). As a result, large ANE has been observed in antiferromagnetic Weyl semimetal $Mn_3Sn$ at order-of-magnitudes lower magnetization comparing to conventional ferromagnetic metals (**Figure 6b**), moreover very large ANE signals have also been observed in ferromagnetic Weyl semimetals $Co_3Sn_2S_2$ and $Co_2MnGa$ (**Figure 6c**).

**Summary and outlook**

In this paper, we introduce how longitudinal and transverse thermal transport may be applied to unveil the key features in a few quantum materials, such as QSLs and superconductors. In these examples, electron correlation plays a crucial role in forming emergent phenomena, with the additional possibility to carry nontrivial topology. Then, correlation and topology can be considered as two pillars that support thermal transport measurements: Correlation can lead to the emergence of new quasiparticles, while topology can lead to exotic boundary states or giant response, as the case of Majorana edge modes in Kitaev QSLs or the giant Nernst effect in topological Weyl semimetals. In light of this, thermal transport may serve as a less-used but powerful alternative to electrical transport when correlation or topology is of interest. Still, there are other areas where thermal transport is intriguing but has to be left out, such as quantization of thermal conductance (*139-*

*141*) and breakdown of Wiedemann-Franz law (*42, 142-144*). In practice, instead of measuring a single value of $k$ (or similarly thermopower $S$), the measurements can be significantly corroborated by performing a temperature dependence $k(T)$, which enables a direct comparison with theory. A further angular degree of freedom $\phi$ with measured $k(T, \phi)$ enables the probe of anisotropy, which has been applied in superconductors but can be applied in a broader type of materials. Moreover, since many correlation effects and their resulting electronic and magnetic orderings are tunable by an external magnetic field $H$, a further field-tunable thermal conductivity $k(T, H, \phi)$ can offer a powerful 3D phase diagram to explore the exotic emergent quasiparticle properties in a wide range of quantum materials.

**Acknowledgments**

M.L. acknowledges support from U.S. DOE BES Award No. DE-SC0020148. G.C. acknowledges the support from U.S. DOE BES Award No. DE-FG02-02ER45977 and the support from ARO MURI (Grant No. W911NF-19-1-0279) via U. Michigan.

**References**

1. The rise of quantum materials. *Nat Phys* **12**, 105-105 (2016).
2. P. W. Anderson, *The theory of superconductivity in the high-Tc cuprates*. Princeton series in physics (Princeton University Press, Princeton, N.J., 1997), pp. vii, 446 p.
3. J. R. Schrieffer, *Theory of superconductivity*. Advanced book classics (Advanced Book Program, Perseus Books, Reading, Mass., 1999), pp. xviii, 332 p.
4. J. Orenstein, A. J. Millis, Advances in the Physics of High-Temperature Superconductivity. *Science* **288**, 468 (2000).
5. A. Damascelli, Z. Hussain, Z.-X. Shen, Angle-resolved photoemission studies of the cuprate superconductors. *Rev Mod Phys* **75**, 473-541 (2003).
6. P. A. Lee, N. Nagaosa, X.-G. Wen, Doping a Mott insulator: Physics of high-temperature superconductivity. *Rev Mod Phys* **78**, 17-85 (2006).
7. K. v. Klitzing, G. Dorda, M. Pepper, New Method for High-Accuracy Determination of the Fine-Structure Constant Based on Quantized Hall Resistance. *Phys Rev Lett* **45**, 494-497 (1980).

8. D. R. Yennie, Integral quantum Hall effect for nonspecialists. *Rev Mod Phys* **59**, 781-824 (1987).
9. B. A. Bernevig, T. L. Hughes, S. C. Zhang, Quantum spin Hall effect and topological phase transition in HgTe quantum wells. *Science* **314**, 1757-1761 (2006).
10. B. A. Bernevig, S. C. Zhang, Quantum spin Hall effect. *Phys Rev Lett* **96**, 106802 (2006).
11. M. Konig *et al.*, Quantum spin hall insulator state in HgTe quantum wells. *Science* **318**, 766-770 (2007).
12. R. Yu *et al.*, Quantized anomalous Hall effect in magnetic topological insulators. *Science* **329**, 61-64 (2010).
13. C. Z. Chang *et al.*, Experimental observation of the quantum anomalous Hall effect in a magnetic topological insulator. *Science* **340**, 167-170 (2013).
14. Z. F. Ezawa, *Quantum Hall effects : recent theoretical and experimental developments*. (World Scientific, New Jersey, ed. Third edition., 2013), pp. 891 pages.
15. C. Z. Chang, M. Li, Quantum anomalous Hall effect in time-reversal-symmetry breaking topological insulators. *Journal of physics. Condensed matter : an Institute of Physics journal* **28**, 123002 (2016).
16. M. Z. Hasan, C. L. Kane, Colloquium: Topological insulators. *Rev Mod Phys* **82**, 3045-3067 (2010).
17. X.-L. Qi, S.-C. Zhang, Topological insulators and superconductors. *Rev Mod Phys* **83**, 1057-1110 (2011).
18. J. Wang, S. C. Zhang, Topological states of condensed matter. *Nat Mater* **16**, 1062-1067 (2017).
19. A. A. Burkov, Topological semimetals. *Nat Mater* **15**, 1145-1148 (2016).
20. B. Yan, C. Felser, Topological Materials: Weyl Semimetals. *Annual Review of Condensed Matter Physics* **8**, 337-354 (2017).
21. N. P. Armitage, E. J. Mele, A. Vishwanath, Weyl and Dirac semimetals in three-dimensional solids. *Rev Mod Phys* **90**, 015001 (2018).
22. K. Manna, Y. Sun, L. Muechler, J. Kübler, C. Felser, Heusler, Weyl and Berry. *Nature Reviews Materials* **3**, 244-256 (2018).
23. M. Neupane *et al.*, Observation of a three-dimensional topological Dirac semimetal phase in high-mobility $Cd_3As_2$. *Nat Commun* **5**, 3786 (2014).
24. S. M. Huang *et al.*, A Weyl Fermion semimetal with surface Fermi arcs in the transition metal monopnictide TaAs class. *Nat Commun* **6**, 7373 (2015).
25. S.-Y. Xu *et al.*, Discovery of a Weyl fermion semimetal and topological Fermi arcs. *Science* **349**, 613 (2015).
26. S.-Y. Xu *et al.*, Observation of Fermi arc surface states in a topological metal. *Science* **347**, 294 (2015).
27. T. Ideue, T. Kurumaji, S. Ishiwata, Y. Tokura, Giant thermal Hall effect in multiferroics. *Nat Mater* **16**, 797-802 (2017).
28. D. Khomskii, Classifying multiferroics: Mechanisms and effects. *Physics* **2**, (2009).

29. H. T. Diep, *Frustrated spin systems*. (World Scientific, Hackensack New Jersey, ed. 2nd edition., 2013), pp. xxv, 617 pages.
30. L. Balents, Spin liquids in frustrated magnets. *Nature* **464**, 199-208 (2010).
31. S. H. Lee *et al.*, Emergent excitations in a geometrically frustrated magnet. *Nature* **418**, 856-858 (2002).
32. S. T. Bramwell, M. J. P. Gingras, Spin Ice State in Frustrated Magnetic Pyrochlore Materials. *Science* **294**, 1495 (2001).
33. N. Read, S. Sachdev, Large-N expansion for frustrated quantum antiferromagnets. *Phys Rev Lett* **66**, 1773-1776 (1991).
34. F. Bonaccorso, Z. Sun, T. Hasan, a. C. Ferrari, Graphene photonics and optoelectronics. *Nature Photonics* **4**, 611-622 (2010).
35. C. W. J. Beenakker, Colloquium: Andreev reflection and Klein tunneling in graphene. *Rev Mod Phys* **80**, 1337-1354 (2008).
36. A. K. Geim, K. S. Novoselov, The rise of graphene. *Nat Mater* **6**, 183-191 (2007).
37. K. S. Novoselov *et al.*, Two-dimensional gas of massless Dirac fermions in graphene. *Nature* **438**, 197-200 (2005).
38. C. L. Kane, E. J. Mele, Quantum spin Hall effect in graphene. *Phys Rev Lett* **95**, 226801 (2005).
39. Q. H. Wang, K. Kalantar-Zadeh, A. Kis, J. N. Coleman, M. S. Strano, Electronics and optoelectronics of two-dimensional transition metal dichalcogenides. *Nat Nanotechnol* **7**, 699-712 (2012).
40. K. F. Mak, J. Shan, Photonics and optoelectronics of 2D semiconductor transition metal dichalcogenides. *Nature Photonics* **10**, 216-226 (2016).
41. C. Kittel, *Introduction to solid state physics*. (Wiley, Hoboken, NJ, ed. 8th, 2005), pp. xix, 680 p.
42. S. Lee *et al.*, Anomalously low electronic thermal conductivity in metallic vanadium dioxide. *Science* **355**, 371 (2017).
43. H. J. Goldsmid, *Introduction to thermoelectricity*. Springer series in materials science, (Springer, Heidelberg ; New York, 2010), pp. xvi, 242 p.
44. Y. Kasahara *et al.*, Majorana quantization and half-integer thermal quantum Hall effect in a Kitaev spin liquid. *Nature* **559**, 227-231 (2018).
45. J. Nasu, M. Udagawa, Y. Motome, Thermal fractionalization of quantum spins in a Kitaev model: Temperature-linear specific heat and coherent transport of Majorana fermions. *Physical Review B* **92**, (2015).
46. Z. Wang, X.-L. Qi, S.-C. Zhang, Topological field theory and thermal responses of interacting topological superconductors. *Physical Review B* **84**, (2011).
47. M. Banerjee *et al.*, Observed quantization of anyonic heat flow. *Nature* **545**, 75-79 (2017).
48. M. Banerjee *et al.*, Observation of half-integer thermal Hall conductance. *Nature* **559**, 205-210 (2018).
49. K. Behnia, H. Aubin, Nernst effect in metals and superconductors: a review of concepts and experiments. *Rep Prog Phys* **79**, 046502 (2016).

50. M. Ikhlas *et al.*, Large anomalous Nernst effect at room temperature in a chiral antiferromagnet. *Nat Phys* **13**, 1085-1090 (2017).
51. F. Caglieris *et al.*, Anomalous Nernst effect and field-induced Lifshitz transition in the Weyl semimetals TaP and TaAs. *Physical Review B* **98**, (2018).
52. S. J. Watzman *et al.*, Dirac dispersion generates unusually large Nernst effect in Weyl semimetals. *Physical Review B* **97**, (2018).
53. L. Savary, L. Balents, Quantum spin liquids: a review. *Rep Prog Phys* **80**, 016502 (2016).
54. Y. Zhou, K. Kanoda, T.-K. Ng, Quantum spin liquid states. *Rev Mod Phys* **89**, (2017).
55. Y. Shimizu, K. Miyagawa, K. Kanoda, M. Maesato, G. Saito, Spin liquid state in an organic Mott insulator with a triangular lattice. *Phys Rev Lett* **91**, 107001 (2003).
56. M. Yamashita *et al.*, Highly Mobile Gapless Excitations in a Two-Dimensional Candidate Quantum Spin Liquid. *Science* **328**, 1246-1248 (2010).
57. Y. Shen *et al.*, Evidence for a spinon Fermi surface in a triangular-lattice quantum-spin-liquid candidate. *Nature*, (2016).
58. Y. Xu *et al.*, Absence of Magnetic Thermal Conductivity in the Quantum Spin-Liquid Candidate YbMgGaO$_4$. *Phys Rev Lett* **117**, 267202 (2016).
59. T. H. Han *et al.*, Fractionalized excitations in the spin-liquid state of a kagome-lattice antiferromagnet. *Nature* **492**, 406-410 (2012).
60. J. S. Helton *et al.*, Spin dynamics of the spin-1/2 kagome lattice antiferromagnet ZnCu3(OH)6Cl2. *Phys Rev Lett* **98**, 107204 (2007).
61. J. S. Gardner, M. J. P. Gingras, J. E. Greedan, Magnetic pyrochlore oxides. *Rev Mod Phys* **82**, 53-107 (2010).
62. Y. Okamoto, M. Nohara, H. Aruga-Katori, H. Takagi, Spin-liquid state in the S=1/2 hyperkagome antiferromagnet Na4Ir3O8. *Phys Rev Lett* **99**, 137207 (2007).
63. A. Kitaev, Anyons in an exactly solved model and beyond. *Annals of Physics* **321**, 2-111 (2006).
64. Y. Singh, P. Gegenwart, Antiferromagnetic Mott insulating state in single crystals of the honeycomb lattice material gamma-Na2IrO3. *Physical Review B* **82**, (2010).
65. A. Banerjee *et al.*, Excitations in the field-induced quantum spin liquid state of α-RuCl3. *npj Quantum Materials* **3**, (2018).
66. A. Banerjee *et al.*, Neutron scattering in the proximate quantum spin liquid alpha-RuCl3. *Science* **356**, 1055-1058 (2017).
67. S.-H. Do *et al.*, Majorana fermions in the Kitaev quantum spin system α-RuCl3. *Nat Phys* **13**, 1079-1084 (2017).
68. A. Biffin *et al.*, Noncoplanar and Counterrotating Incommensurate Magnetic Order Stabilized by Kitaev Interactions in gamma-NaIrO$_3$. *Phys Rev Lett* **113**, 197201 (2014).
69. Z. Ma *et al.*, Recent progress on magnetic-field studies on quantum-spin-liquid candidates. *Chinese Physics B* **27**, 106101 (2018).

70. J. Wen, S.-L. Yu, S. Li, W. Yu, J.-X. Li, Experimental identification of quantum spin liquids. *npj Quantum Materials* **4**, (2019).
71. S. M. Winter *et al.*, Models and materials for generalized Kitaev magnetism. *Journal of physics. Condensed matter : an Institute of Physics journal* **29**, 493002 (2017).
72. R. Hentrich *et al.*, Unusual Phonon Heat Transport in alpha-RuCl_{3}: Strong Spin-Phonon Scattering and Field-Induced Spin Gap. *Phys Rev Lett* **120**, 117204 (2018).
73. Y. J. Yu *et al.*, Ultralow-Temperature Thermal Conductivity of the Kitaev Honeycomb Magnet alpha-RuCl_{3} across the Field-Induced Phase Transition. *Phys Rev Lett* **120**, 067202 (2018).
74. J. Cookmeyer, J. E. Moore, Spin-wave analysis of the low-temperature thermal Hall effect in the candidate Kitaev spin liquid $\alpha-$RuCl3. *Physical Review B* **98**, (2018).
75. H. Doki *et al.*, Spin Thermal Hall Conductivity of a Kagome Antiferromagnet. *Phys Rev Lett* **121**, 097203 (2018).
76. Y. Kasahara *et al.*, Unusual Thermal Hall Effect in a Kitaev Spin Liquid Candidate $\alpha-$RuCl3. *Phys Rev Lett* **120**, (2018).
77. M. Ye, G. B. Halasz, L. Savary, L. Balents, Quantization of the Thermal Hall Conductivity at Small Hall Angles. *Phys Rev Lett* **121**, 147201 (2018).
78. T. Isono *et al.*, Spin-lattice decoupling in a triangular-lattice quantum spin liquid. *Nat Commun* **9**, 1509 (2018).
79. J. M. Ni *et al.*, Ultralow-temperature heat transport in the quantum spin liquid candidate Ca10Cr7O28 with a bilayer kagome lattice. *Physical Review B* **97**, (2018).
80. L. Clark *et al.*, Two-dimensional spin liquid behaviour in the triangular-honeycomb antiferromagnet TbInO3. *Nat Phys* **15**, 262-268 (2019).
81. R. Samajdar, S. Chatterjee, S. Sachdev, M. S. Scheurer, Thermal Hall effect in square-lattice spin liquids: A Schwinger boson mean-field study. *Physical Review B* **99**, (2019).
82. Y. H. Gao, C. Hickey, T. Xiang, S. Trebst, G. Chen, Thermal Hall signatures of non-Kitaev spin liquids in honeycomb Kitaev materials. *Physical Review Research* **1**, (2019).
83. A. Pidatella, A. Metavitsiadis, W. Brenig, Heat transport in the anisotropic Kitaev spin liquid. *Physical Review B* **99**, (2019).
84. A. Metavitsiadis, C. Psaroudaki, W. Brenig, Spin liquid fingerprints in the thermal transport of a Kitaev-Heisenberg ladder. *Physical Review B* **99**, (2019).
85. Z. Ma *et al.*, Spin-Glass Ground State in a Triangular-Lattice Compound YbZnGaO$_4$. *Phys Rev Lett* **120**, 087201 (2018).
86. J. M. Ni *et al.*, Absence of Magnetic Thermal Conductivity in the Quantum Spin Liquid Candidate EtMe3Sb[Pd(dmit)2]2. *Phys Rev Lett* **123**, (2019).

87. P. Bourgeois-Hope *et al.*, Thermal Conductivity of the Quantum Spin Liquid Candidate EtMe3Sb[Pd(dmit)2]2: No Evidence of Mobile Gapless Excitations. *Physical Review X* **9**,   (2019).
88. M. Yamashita *et al.*, Thermal-transport measurements in a quantum spin-liquid state of the frustrated triangular magnet κ-(BEDT-TTF)2Cu2(CN)3. *Nat Phys* **5**, 44-47 (2008).
89. M. Hirschberger, R. Chisnell, Y. S. Lee, N. P. Ong, Thermal Hall Effect of Spin Excitations in a Kagome Magnet. *Phys Rev Lett* **115**, 106603 (2015).
90. H. Katsura, N. Nagaosa, P. A. Lee, Theory of the thermal Hall effect in quantum magnets. *Phys Rev Lett* **104**, 066403 (2010).
91. Y. Matsuda, K. Izawa, I. Vekhter, Nodal structure of unconventional superconductors probed by angle resolved thermal transport measurements. *Journal of Physics: Condensed Matter* **18**, R705-R752 (2006).
92. C. Uher, Y. Q. Liu, J. F. Whitaker, The Peak in the Thermal-Conductivity of Cu-O Superconductors - Electronic or Phononic Origin. *Journal of Superconductivity* **7**, 323-329 (1994).
93. C. Proust, E. Boaknin, R. W. Hill, L. Taillefer, A. P. Mackenzie, Heat transport in a strongly overdoped cuprate: Fermi liquid and a pure d-wave BCS superconductor. *Phys Rev Lett* **89**, 147003 (2002).
94. S. J. Hagen, Z. Z. Wang, N. P. Ong, Anisotropy of the thermal conductivity of YBa2Cu3O7-y. *Phys Rev B Condens Matter* **40**, 9389-9392 (1989).
95. M. Matsukawa, T. Mizukoshi, K. Noto, Y. Shiohara, In-plane and out-of-plane thermal conductivity of a large single crystal of YBa2Cu3O7-x. *Physical Review B* **53**, R6034-R6037 (1996).
96. G. K. White, S. J. Collocott, R. Driver, R. B. Roberts, A. M. Stewart, Thermal-Properties of High-Tc Superconductors. *J Phys C Solid State* **21**, L631-L637 (1988).
97. D. Varshney, K. K. Choudhary, R. K. Singh, Analysis of in-plane thermal conductivity anomalies in YBa2Cu3O7-delta cuprate superconductors. *New J Phys* **5**,   (2003).
98. M. B. Salamon, F. Yu, V. N. Kopylov, The Field-Dependence of the Thermal-Conductivity - Evidence for Nodes in the Gap. *Journal of Superconductivity* **8**, 449-452 (1995).
99. M. I. Flik, C. L. Tien, Size Effect on the Thermal-Conductivity of High-Tc Thin-Film Superconductors. *J Heat Trans-T Asme* **112**, 872-881 (1990).
100. L. Tewordt, T. Wölkhausen, Theory of phonon thermal conductivity for strong-coupling s- and d-wave pairing in high Tc superconductors. *Solid State Communications* **75**, 515-519 (1990).
101. M. J. Graf, S. K. Yip, J. A. Sauls, D. Rainer, Electronic thermal conductivity and the Wiedemann-Franz law for unconventional superconductors. *Physical Review B* **53**, 15147-15161 (1996).
102. P. J. Hirschfeld, W. O. Putikka, Theory of thermal conductivity in YBa2Cu3O7-delta. *Phys Rev Lett* **77**, 3909-3912 (1996).

103. A. S. Alexandrov, N. F. Mott, Thermal transport in a charged Bose gas and in high-Tc oxides. *Phys Rev Lett* **71**, 1075-1078 (1993).
104. S. Wermbter, L. Tewordt, Density of States, Nuclear-Relaxation Rate, and Thermal-Conductivity in High-Tc Superconductors. *Physical Review B* **44**, 9524-9530 (1991).
105. K. E. Goodson, M. I. Flik, Electron and Phonon Thermal Conduction in Epitaxial High-Tc Superconducting Films. *J Heat Trans-T Asme* **115**, 17-25 (1993).
106. T. Muranaka, J. Akimitsu, M. Sera, Thermal transport properties of MgB2. *Physical Review B* **64**,  (2001).
107. A. V. Sologubenko, J. Jun, S. M. Kazakov, J. Karpinski, H. R. Ott, Thermal conductivity of single-crystalline MgB2. *Physical Review B* **66**, (2002).
108. M. Putti *et al.*, Thermal conductivity of MgB2 in the superconducting state. *Physical Review B* **67**,  (2003).
109. R. Movshovich *et al.*, Unconventional superconductivity in CeIrIn5 and CeCoIn5: specific heat and thermal conductivity studies. *Phys Rev Lett* **86**, 5152-5155 (2001).
110. K. Izawa *et al.*, Multiple superconducting phases in new heavy fermion superconductor PrOs4Sb12. *Phys Rev Lett* **90**,  (2003).
111. G. Seyfarth *et al.*, Superconducting PrOs4Sb12: a thermal conductivity study. *Phys Rev Lett* **97**, 236403 (2006).
112. J. P. Reid *et al.*, Universal heat conduction in the iron arsenide superconductor KFe2As2: evidence of a d-wave state. *Phys Rev Lett* **109**, 087001 (2012).
113. J. P. Reid *et al.*, Nodes in the gap structure of the iron arsenide superconductor Ba(Fe1−xCox)2As2 from c-axis heat transport measurements. *Physical Review B* **82**,  (2010).
114. M. A. Tanatar *et al.*, Isotropic three-dimensional gap in the iron arsenide superconductor LiFeAs from directional heat transport measurements. *Physical Review B* **84**,  (2011).
115. Z. Zhang *et al.*, Heat transport in RbFe2As2 single crystals: Evidence for nodal superconducting gap. *Physical Review B* **91**,  (2015).
116. Y. Machida *et al.*, Possible Sign-Reversing-Wave Superconductivity in Co-Doped BaFe2As2 Proved by Thermal Transport Measurements. *J Phys Soc Jpn* **78**, 073705 (2009).
117. G. Chen, *Nanoscale energy transport and conversion : a parallel treatment of electrons, molecules, phonons, and photons*. MIT-Pappalardo series in mechanical engineering (Oxford University Press, Oxford ; New York, 2005), pp. xxiii, 531 p.
118. M. Ausloos, M. Houssa, Thermal conductivity of unconventional superconductors: a probe of the order parameter symmetry. *Supercond Sci Tech* **12**, R103-R114 (1999).
119. H. Shakeripour, C. Petrovic, L. Taillefer, Heat transport as a probe of superconducting gap structure. *New J Phys* **11**, 055065 (2009).

120. P. A. Lee, Localized states in a d-wave superconductor. *Phys Rev Lett* **71**, 1887-1890 (1993).
121. I. M. Lifshitz, Anomalies of Electron Characteristics of a Metal in the High Pressure Region *Journal of Experimental and Theoretical Physics Letters* **11**, 1130 (1960).
122. Z. Zhu, H. Yang, B. Fauqué, Y. Kopelevich, K. Behnia, Nernst effect and dimensionality in the quantum limit. *Nat Phys* **6**, 26-29 (2009).
123. D. Xiao, Y. Yao, Z. Fang, Q. Niu, Berry-phase effect in anomalous thermoelectric transport. *Phys Rev Lett* **97**, 026603 (2006).
124. R. Lundgren, P. Laurell, G. A. Fiete, Thermoelectric properties of Weyl and Dirac semimetals. *Physical Review B* **90**, 165115 (2014).
125. G. Sharma, P. Goswami, S. Tewari, Nernst and magnetothermal conductivity in a lattice model of Weyl fermions. *Physical Review B* **93**, (2016).
126. M. N. Chernodub, A. Cortijo, M. A. H. Vozmediano, Generation of a Nernst Current from the Conformal Anomaly in Dirac and Weyl Semimetals. *Phys Rev Lett* **120**, 206601 (2018).
127. J. Noky, J. Gayles, C. Felser, Y. Sun, Strong anomalous Nernst effect in collinear magnetic Weyl semimetals without net magnetic moments. *Physical Review B* **97**, (2018).
128. S. Saha, S. Tewari, Anomalous Nernst effect in type-II Weyl semimetals. *The European Physical Journal B* **91**, (2018).
129. K. Das, A. Agarwal, Berry curvature induced thermopower in type-I and type-II Weyl semimetals. *Physical Review B* **100**, (2019).
130. V. Kozii, B. Skinner, L. Fu, Thermoelectric Hall conductivity and figure of merit in Dirac/Weyl materials. *eprint arXiv:1902.10123*, arXiv:1902.10123 (2019).
131. R. C. McKay, T. M. McCormick, N. Trivedi, Nernst thermopower of time-reversal breaking type-II Weyl semimetals. *Physical Review B* **99**, (2019).
132. T. Liang *et al.*, Anomalous Nernst Effect in the Dirac Semimetal $Cd_3As_2$. *Phys Rev Lett* **118**, 136601 (2017).
133. G. Sharma, C. Moore, S. Saha, S. Tewari, Nernst effect in Dirac and inversion-asymmetric Weyl semimetals. *Physical Review B* **96**, (2017).
134. C. Fu *et al.*, Large Nernst power factor over a broad temperature range in polycrystalline Weyl semimetal NbP. *Energy & Environmental Science* **11**, 2813-2820 (2018).
135. K. G. Rana *et al.*, Thermopower and Unconventional Nernst Effect in the Predicted Type-II Weyl Semimetal WTe2. *Nano Lett* **18**, 6591-6596 (2018).
136. H. Reichlova *et al.*, Large anomalous Nernst effect in thin films of the Weyl semimetal Co2MnGa. *Appl Phys Lett* **113**, 212405 (2018).
137. S. N. Guin *et al.*, Zero-Field Nernst Effect in a Ferromagnetic Kagome-Lattice Weyl-Semimetal Co3 Sn2 S2. *Adv Mater*, e1806622 (2019).
138. C. Wuttke *et al.*, Berry curvature unravelled by the anomalous Nernst effect in Mn3Ge. *Physical Review B* **100**, (2019).

139. K. Schwab, E. A. Henriksen, J. M. Worlock, M. L. Roukes, Measurement of the quantum of thermal conductance. *Nature* **404**, 974-977 (2000).
140. D. E. Angelescu, M. C. Cross, M. L. Roukes, Heat transport in mesoscopic systems. *Superlattices and Microstructures* **23**, 673-689 (1998).
141. L. G. C. Rego, G. Kirczenow, Quantized Thermal Conductance of Dielectric Quantum Wires. *Phys Rev Lett* **81**, 232-235 (1998).
142. M. A. Tanatar, J. Paglione, C. Petrovic, L. Taillefer, Anisotropic violation of the Wiedemann-Franz law at a quantum critical point. *Science* **316**, 1320-1322 (2007).
143. H. Pfau *et al.*, Thermal and electrical transport across a magnetic quantum critical point. *Nature* **484**, 493-497 (2012).
144. J. Crossno *et al.*, Observation of the Dirac fluid and the breakdown of the Wiedemann-Franz law in graphene. *Science* **351**, 1058-1061 (2016).
145. I. Santoso, Van der Waals-Zeeman Institute (WZI), Amsterdam (2008).

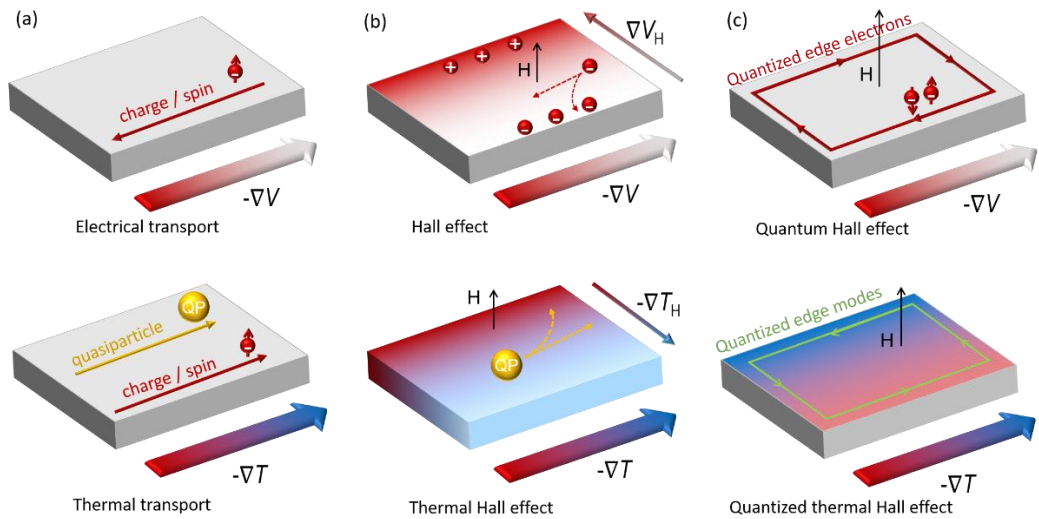

**Figure 1.** (a) Longitudinal electrical (top) and thermal (bottom) transport. Comparing to electrical transport using voltage drop $-\nabla V$ as perturbation, thermal transport uses temperature drop $-\nabla T$. Thermal transport is also sensitive to mobile quasiparticles without carrying electrical charge. (b) (Electrical) Hall effect (top) and thermal Hall (bottom) transport, where the measured response is the transverse voltage difference $\nabla V_H$ and transverse temperature difference $\nabla T_H$, respectively. (c) Just like quantum Hall effect with chiral edge electronic states (top), the quantized thermal Hall effect can be used to probe charge-neutral edge states (bottom).

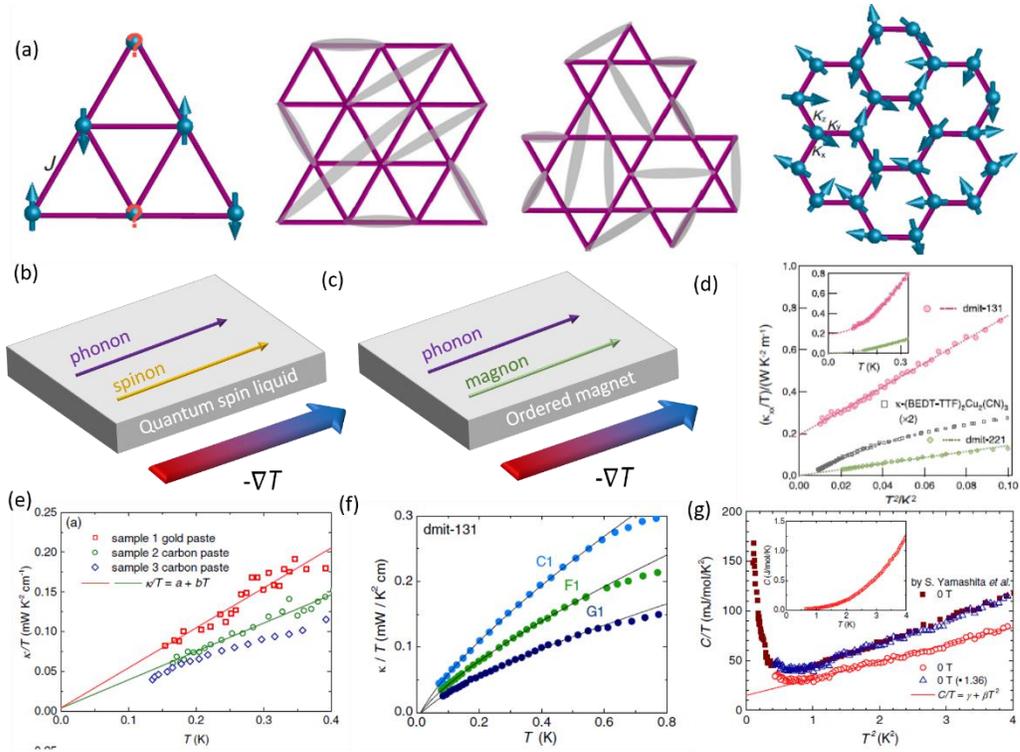

**Figure 2.** (a) Geometrical frustration (left three figures are triangular, triangular and kagome) and Kitaev (right) interaction toward a QSL state. (b) The spinons in QSL and (c) the magnons in ordered magnets in addition to phonons as elementary excitations. (d) The observation of linear residual term of thermal conductivity (pink curve) and the absence of linear residual term (e, f) in QSL candidate EtMe$_3$Sb[Pd(dmit)$_2$]$_2$. (g) The heat capacity shows the existence of the linear residual term (red curve) even if it is absent in thermal transport. Figure (a) adapted from (*70*), (d) adapted from (*56*), (e,g) adapted from (*86*) and (f) adapted from (*87*).

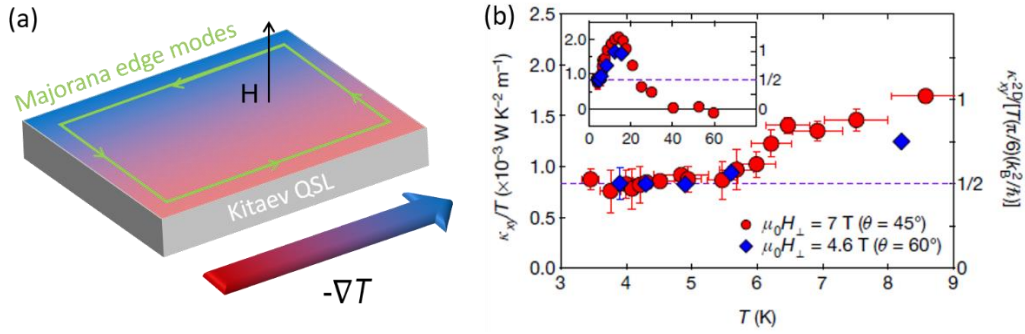

**Figure 3.** (a) The schematics of Majorana edge modes in a Kitaev QSL and the resulted (b) half-integer quantization of thermal Hall effect. Figure (b) adapted from (*44*).

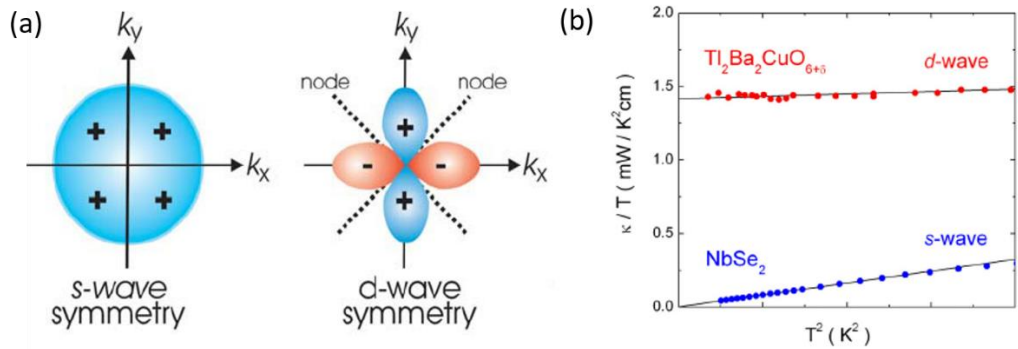

**Figure 4.** (a) *s*-wave vs *d*-wave symmetry of superconducting gap in 2D **k**-space. (b) The example of using the linear residual term in *k*/*T* to distinguish a *d*-wave superconductor and an *s*-wave superconductor. Figure (a) adapted from (*145*), figure (b) adapted from (*119*).

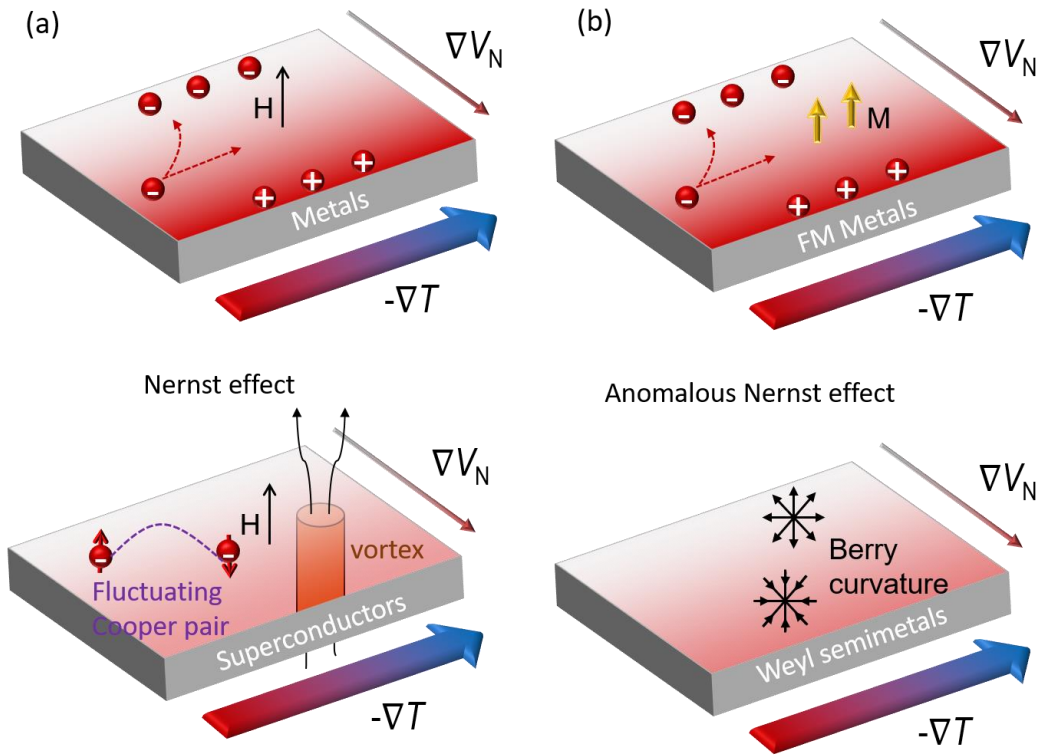

**Figure 5.** (a) Nernst effect in metals (top) and superconductors (bottom). (b) The anomalous Nernst effect in ferromagnetic (FM) metals (top) and Weyl semimetals (bottom) with nontrivial Berry curvature.

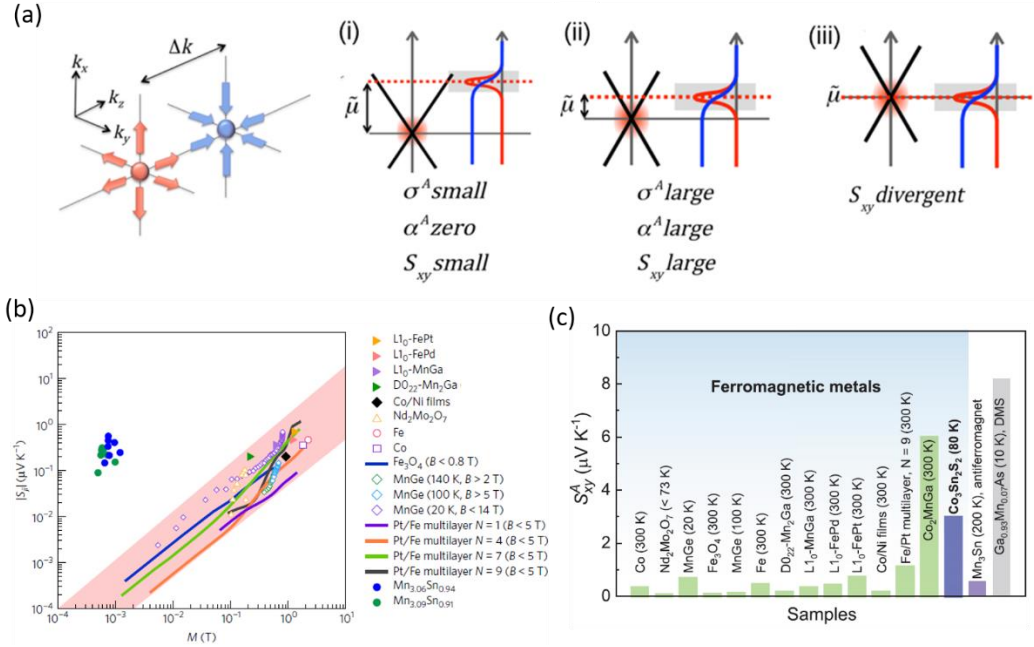

**Figure 6.** (a) (Left) The paired Weyl nodes as source (red arrows) and sink (blue arrows) of Berry curvature, and the sensitivity of Nernst effect to the relative shift between chemical potential $\tilde{\mu}$ and Weyl node. Cases i-iii are the situations where (i) $\tilde{\mu}$ is far away from Weyl node, (ii) $\tilde{\mu}$ is close to Weyl node (ii), and $\tilde{\mu}$ is located right at the Weyl node (iii) with divergent Berry curvature. (b, c) The anomalous Nernst effect expressed in terms of transverse thermopower in antiferromagnetic (b) and ferromagnetic (c) Weyl semimetals comparing to typical ferromagnetic metals, where the large magnitudes are attributed to nontrivial Berry curvature. Figure (a) adapted from (*51*), figure (b) adapted from (*50*), figure (c) adapted from (*137*).